\pdfoutput=1
\RequirePackage{ifpdf}
\ifpdf 
\documentclass[pdftex]{sigma}
\else
\documentclass{sigma}
\fi

\numberwithin{equation}{section}

\begin{document}
\allowdisplaybreaks

\renewcommand{\thefootnote}{}

\newcommand{\arXivNumber}{2301.03021}

\renewcommand{\PaperNumber}{051}

\FirstPageHeading

\ShortArticleName{A Skyrme Model with Novel Chiral Symmetry Breaking}

\ArticleName{A Skyrme Model with Novel Chiral Symmetry\\ Breaking\footnote{This paper is a~contribution to the Special Issue on Topological Solitons as Particles. The~full collection is available at \href{https://www.emis.de/journals/SIGMA/topological-solitons.html}{https://www.emis.de/journals/SIGMA/topological-solitons.html}}}

\Author{Paul SUTCLIFFE}

\AuthorNameForHeading{P.~Sutcliffe}

\Address{Department of Mathematical Sciences, Durham University, Durham DH1 3LE, UK}
\Email{\href{mailto:p.m.sutcliffe@durham.ac.uk}{p.m.sutcliffe@durham.ac.uk}}

\ArticleDates{Received February 16, 2023, in final form July 21, 2023; Published online July 26, 2023}

\Abstract{An extension of the Skyrme model is presented in which derivative terms are added that break chiral symmetry to isospin symmetry. The theory contains just one new parameter and it reduces to the standard Skyrme model when this symmetry breaking parameter vanishes. The same Faddeev--Bogomolny energy bound applies for all parameter values, but the parameter can be tuned so that the energy of the single Skyrmion is much closer to the bound than in the standard Skyrme model. Applying the rational map approximation to multi-Skyrmions suggests that, for a suitable value of the symmetry breaking parameter, binding energies in this theory may be significantly more realistic than in the standard Skyrme model.}

\Classification{35C08; 35Q51}

\renewcommand{\thefootnote}{\arabic{footnote}}
\setcounter{footnote}{0}

\section{Introduction}

Skyrmions are topological solitons that model nuclei within a nonlinear theory of pions~\cite{Manton:2022fcb}. There have been many developments on Skyrmions since Skyrme's original work~\cite{Skyrme:1962vh}, and during the last decade or so some attention has been directed towards modifications of the Skyrme model that bring the energy of a single Skyrmion closer to the Faddeev--Bogomolny energy bound~\cite{Faddeev:1976pg}. The motivation for this is to address the issue that Skyrmions are too tightly bound in comparison to nuclei. If the energy of a single Skyrmion is close to the bound then clearly this places a tight restriction on multi-Skyrmion binding energies. However, in the standard Skyrme model the single Skyrmion energy exceeds the bound by around $23\%$ and this leaves plenty of room for large binding energies, well in excess of those of nuclei, that are no greater than around $1\%.$

Some examples of modifications to the standard Skyrme model to address this problem include the BPS sextic model~\cite{Adam:2013wya,Adam:2010fg}, the lightly bound model~\cite{Gillard:2015eia, Harland:2013rxa}, the loosely bound model~\cite{Gudnason:2016mms}, the dielectric model~\cite{Adam:2020iye}, the self-dual model~\cite{Ferreira:2017bsr}, the quasi self-dual model~\cite{Ferreira:2022zkm}, the inclusion of~$\omega$ mesons~\cite{Gudnason:2020arj}, and the addition of $\rho$ mesons via Yang--Mills theory with an extra dimension \cite{Naya:2018kyi,Naya:2018mpt,Sutcliffe:2010et,Sutcliffe:2011ig}. Each theory has its own positive aspects that make it worthy of investigation, together with negative aspects that mean that alternatives are still being sought. The purpose of the present paper is to add a new model to the list of modified theories that bring Skyrmion energies closer to the energy bound. This is achieved by the introduction of derivative terms that break chiral symmetry to isospin symmetry. The additional terms come with one new parameter that can be chosen so that the energy of the single Skyrmion is reduced from $23\%$ to~$7\%$ above the bound. An investigation of multi-Skyrmions within this theory, using the rational map approximation~\cite{Houghton:1997kg}, suggests that more realistic binding energies may indeed be possible in this broken theory.

\section{A Skyrme model from Yang--Mills}

The standard Skyrme model is a nonlinear theory of pions in which the pion fields $(\pi_1,\pi_2,\pi_3)$ are used to construct
the Skyrme field $U\in {\rm SU}(2)$ via
\[
U=\begin{pmatrix}
\sigma +{\rm i}\pi_3 & {\rm i}\pi_1+\pi_2 \\ {\rm i}\pi_1-\pi_2 & \sigma -{\rm i}\pi_3
\end{pmatrix},
\]
where the $\sigma$ field imposes the constraint $\sigma^2+\pi_1^2+\pi_2^2+\pi_3^2=1.$ In Skyrme units, the static energy is given by
\begin{equation}
E_{\rm Skyrme}=-\int
\operatorname{Tr}\bigg\{\frac{1}{2}R_i^2+\frac{1}{16}[R_i,R_j]^2
\bigg\} {\rm d}^3x,
\label{enpion}
\end{equation}
where the $\mathfrak{su}(2)$-valued right currents are $R_i=\partial_iU U^{-1}.$

This theory is invariant under chiral symmetry, given by multiplication of the Skyrme field by an arbitrary constant ${\rm SU}(2)$ matrix on the left and another arbitrary constant ${\rm SU}(2)$ matrix on the right, which together rotates the quartet of the $\sigma$ and pion fields. Finite energy imposes the boundary condition that the Skyrme field tends to a constant ${\rm SU}(2)$ matrix at infinity, taken to be the identity matrix. This breaks chiral symmetry to isospin symmetry, given by conjugation of the Skyrme field by an arbitrary constant ${\rm SU}(2)$ matrix, which rotates the pions fields but leaves the $\sigma$ field unchanged.
Chiral symmetry can also be broken explicitly by the addition of a potential term to the energy~\eqref{enpion}. The most common choice is to add the term $2m^2(1-\sigma)$, which gives the pions a mass $m$ in Skyrme units. However, this study will be restricted to the case of massless pions.

The boundary condition provides a compactification of space, therefore topologically $U$ is a~map between three-spheres with an associated integer $B$, the topological degree, that may be computed as
\[
B=\int \frac{1}{24\pi^2}\varepsilon_{ijk}\operatorname{Tr}(R_iR_kR_j)\, {\rm d}^3x.
\]
This integer is identified with baryon number and is also referred to as the topological charge. Skyrmions are minimizers of the energy for a given positive value of $B$ and their energies satisfy the Faddeev--Bogomolny energy bound~\cite{Faddeev:1976pg}
\begin{equation}
E_{\rm Skyrme}\ge 12\pi^2 B.
\label{fb}
\end{equation}
As mentioned above, for the single Skyrmion ($B=1$) the energy exceeds this bound by around~$23\%$ and an aim is to modify the theory to reduce this excess.

An interesting perspective on the Skyrme model is obtained via consideration of ${\rm SU}(2)$ Yang--Mills theory with an additional spatial dimension. In four-dimensional Euclidean space the Yang--Mills energy (with a non-standard normalization) is given by
\begin{equation}
E=-\frac{3}{4}\int \operatorname{Tr}(F_{IJ}F_{IJ})\,{\rm d}^4x,
\label{yme}
\end{equation}
where $x_I$ with $I=1,\dots,4$ denote the spatial coordinates
and
$F_{IJ}=\partial_I A_J-\partial_J A_I+[A_I,A_J]$ are the components
of the $\mathfrak{su}(2)$-valued field strength.
The instanton number $N\in\mathbb{Z}$ of the gauge field provides a lower bound on the energy, which for positive $N$ is{\samepage
\begin{equation}
E\ge 12\pi^2 N.
\label{ymbound}
\end{equation}
Unlike the Skyrme model, this bound is attained for all positive $N$, by fields that are self-dual.}

To make contact with the Skyrme model, fix the gauge $A_4=0$ and restrict the remaining three components $A_i$ to have the form
\begin{equation}
A_i=-\frac{1}{2}(1+\phi)R_i,
\label{basic}
\end{equation}
where $\phi=\tanh x_4$ and $R_i$ is the right current of the Skyrme field $U$, that depends upon the remaining three spatial coordinates $(x_1,x_2,x_3).$
Substituting~\eqref{basic} into the Yang--Mills energy~\eqref{yme} and performing the integration over $x_4$ yields precisely the Skyrme energy~\eqref{enpion}. Furthermore, as $U$ is the holonomy of the gauge field along lines parallel to the $x_4$-axis, the baryon number of the Skyrme field is equal to the instanton number, $B=N$~\cite{Atiyah:1989dq}. The Faddeev--Bogomolny bound~\eqref{fb} therefore follows directly from the Yang--Mills bound~\eqref{ymbound}.

From the Yang--Mills perspective, the Skyrmion excess energy above the bound~\eqref{fb} is a~measure of the failure of the restricted form~\eqref{basic} to describe a self-dual instanton. The excess can be reduced by improving~\eqref{basic} through the addition of a term $\phi' V_i$ to the right-hand side. The~$\mathfrak{su}(2)$-valued vector field $V_i(x_1,x_2,x_3)$ physically describes $\rho$ mesons, the lightest of the vector mesons. Performing the integration over $x_4$ in the Yang--Mills energy using the modified restricted form generates a modified Skyrme theory, coupling pions and $\rho$ mesons~\cite{Naya:2018mpt,Sutcliffe:2010et,Sutcliffe:2011ig}. The Faddeev--Bogomolny bound of $12\pi^2 B$ remains intact, due to its continued inheritance from the Yang--Mills bound, but now the single Skyrmion energy is only around $6\%$ above the bound. The symmetries and shapes of the Skyrmions retain the same forms as in the standard Skyrme model, but binding energies are reduced to about one-quarter of their previous values. Moreover, this flattening of the energy landscape means that the addition of a pion mass term results in a~more dramatic change in the structure of Skyrmions, which for $B>4$ have the cluster structure expected for light nuclei~\cite{Naya:2018kyi}.

The above comments highlight the positive aspects of the Skyrme model with pions and~$\rho$ mesons, but there are also some negative aspects. In particular, with this modification the number of independent field components increases from three to twelve, and in combination with the large number of additional terms in the energy, this makes the theory difficult to work with both analytically and numerically. This motivates the work in the following, to try and produce a theory with similar results but without introducing any additional fields to supplement the Skyrme field.

To derive the new variant of the Skyrme model, first define the twist operator $^+$ and its inverse  $^-$  by the following action on the current
\[
R_i^+=\frac{(1-U^{-1})R_i(1-U)}{2(1-\sigma)}, \qquad
R_i^-=\frac{(1-U)R_i(1-U^{-1})}{2(1-\sigma)}.
\]
A quick calculation shows that a  double application of the twist operator to the right current yields the left current,
$R_i^{++}=U^{-1}\partial_iU.$
Note that for the twisted current to be well-defined as $\sigma\to 1$, which of course corresponds to the Skyrme field approaching its vacuum value $U\to 1$, the current must vanish in this limit, $R_i\to 0.$ This condition, that needs to be imposed on the Skyrme field, is the same condition that is imposed in the standard Skyrme model at spatial infinity, by the requirement of finite energy. As Skyrmions in the standard Skyrme model already satisfy this property then it seems that this requirement is not a significant issue.

The new modified Skyrme model is derived by extending the restricted form~\eqref{basic} to
\begin{equation}
A_i=-\frac{1}{2}(1+\phi)R_i+\frac{\chi \phi'}{\sqrt{1-\sigma}}\big(R_i^--R_i^+\big),
\label{ansatz}
\end{equation}
where $\chi$ is a real constant parameter of the theory. The explicit appearance of the $\sigma$ field breaks chiral symmetry to isospin symmetry, in the same way as the inclusion of the pion mass term, which also explicitly includes the $\sigma$ field. The structure of this term is motivated by symmetry considerations and the aim to have a surrogate for the vector field $V_i$ that is constructed from the Skyrme field and brings the restricted form closer to a self-dual field.

Substituting~\eqref{ansatz} into the Yang--Mills energy~\eqref{yme} and performing the integration over $x_4$ yields the following energy for a modified Skyrme model
\begin{align}
  E={}&-\int\operatorname{Tr}\bigg\{
  \frac{1}{2} R_i^2+\frac{1}{16}[ R_i,R_j]^2
  -\frac{\chi}{4}\big[R_i^+,R_j^+\big]H_{ij}
+\frac{4}{5}\chi^3[D_i,D_j]H_{ij}
  +\frac{24}{35}\chi^4[D_i,D_j]^2\nonumber\\
  &\quad{}+\frac{\chi^2}{20}\Big(
 32D_i^2
  +5H_{ij}^2
   -8\big[R_i^+,R_j^+\big][D_i,D_j]
  +\big(
  \big[R_i^+,D_j\big]\!-\!\big[R_j^+,D_i\big]\big)^2
  \Big)
  \bigg\} {\rm d}^3x.
  \label{csb}
\end{align}
For notational convenience, in the above the $\mathfrak{su}(2)$-valued quantities $D_i$ and $H_{ij}$ have been introduced, where
\[
D_i= \frac{R_i-R_i^{++}}{\sqrt{1-\sigma}},
\]
and $H_{ij}$ is the anti-symmetric tensor
\[
H_{ij}=\frac{2[R_i,R_j]+2\big[R_i^{++},R_j^{++}\big]-[D_i,D_j]}{\sqrt{1-\sigma}}
-\frac{D_i \partial_j\sigma -D_j\partial_i\sigma}{1-\sigma}.
\]
If $\chi=0$ then the energy~\eqref{csb} reverts to that of the standard Skyrme model, but for $\chi\ne0$ it is a deformation via derivative terms that break chiral symmetry to isospin symmetry.
The Faddeev--Bogomolny energy bound, $E\ge 12\pi^2 B$, holds for all $\chi$, as it follows from the Yang--Mills energy bound.

To get a feel for the structure of the additional terms it is perhaps useful to note that in the first two terms, that agree with the standard Skyrme model, all currents $R_i$ could be replaced by their twisted versions $R_i^+$, because both these terms are invariant under this replacement. Regarding $D_i$ as a variant of $R_i^+$, and $H_{ij}$ as a variant of $\big[R_i^+,R_j^+\big]$, the structure of the remaining terms follows from the first two terms by making different possible replacements of the twisted currents and their commutators by these variants.

\section{Skyrmions}

\begin{figure}[t]\centering
       \includegraphics[width=0.46\columnwidth]{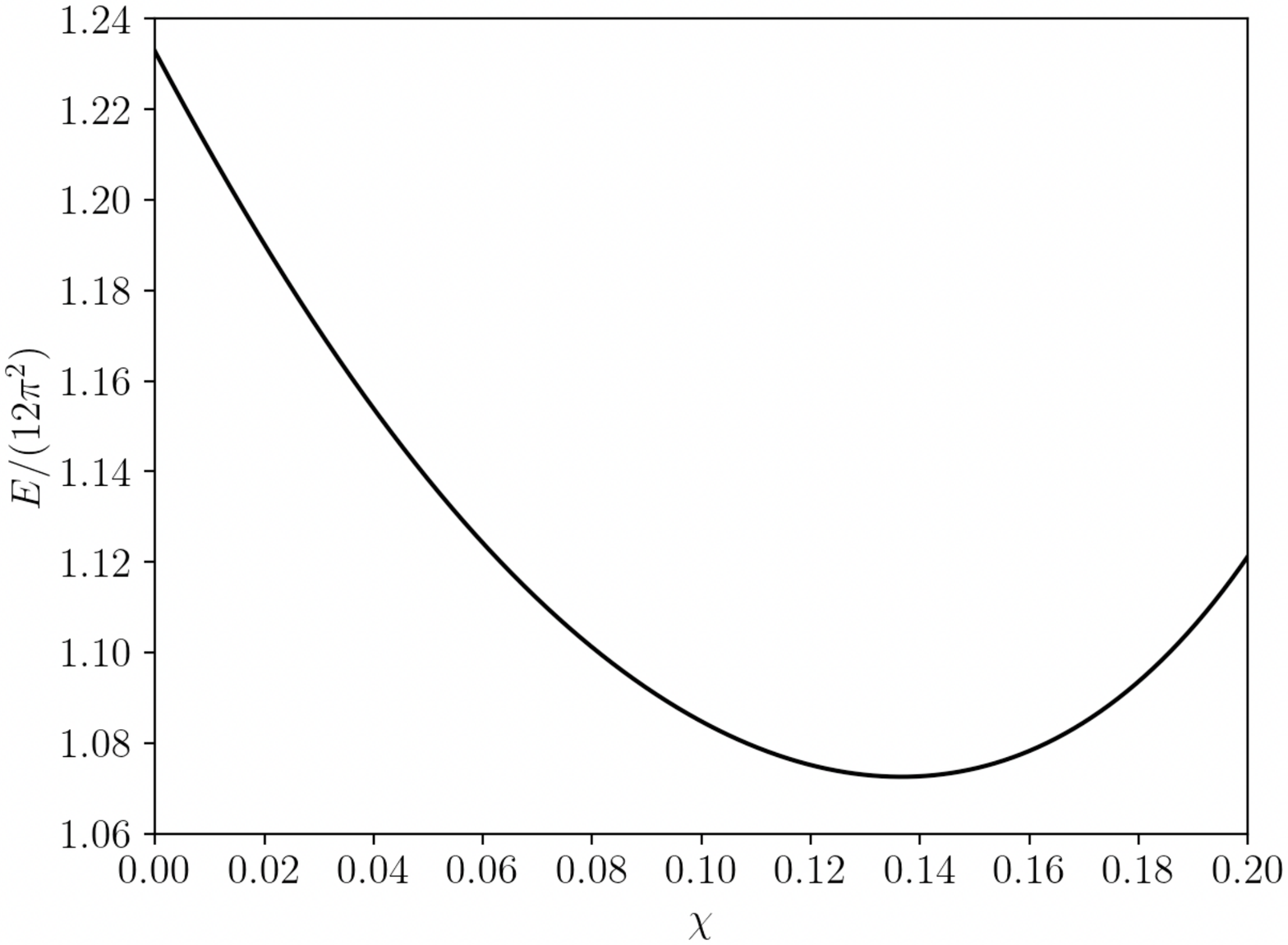} \qquad
      \includegraphics[width=0.46\columnwidth]{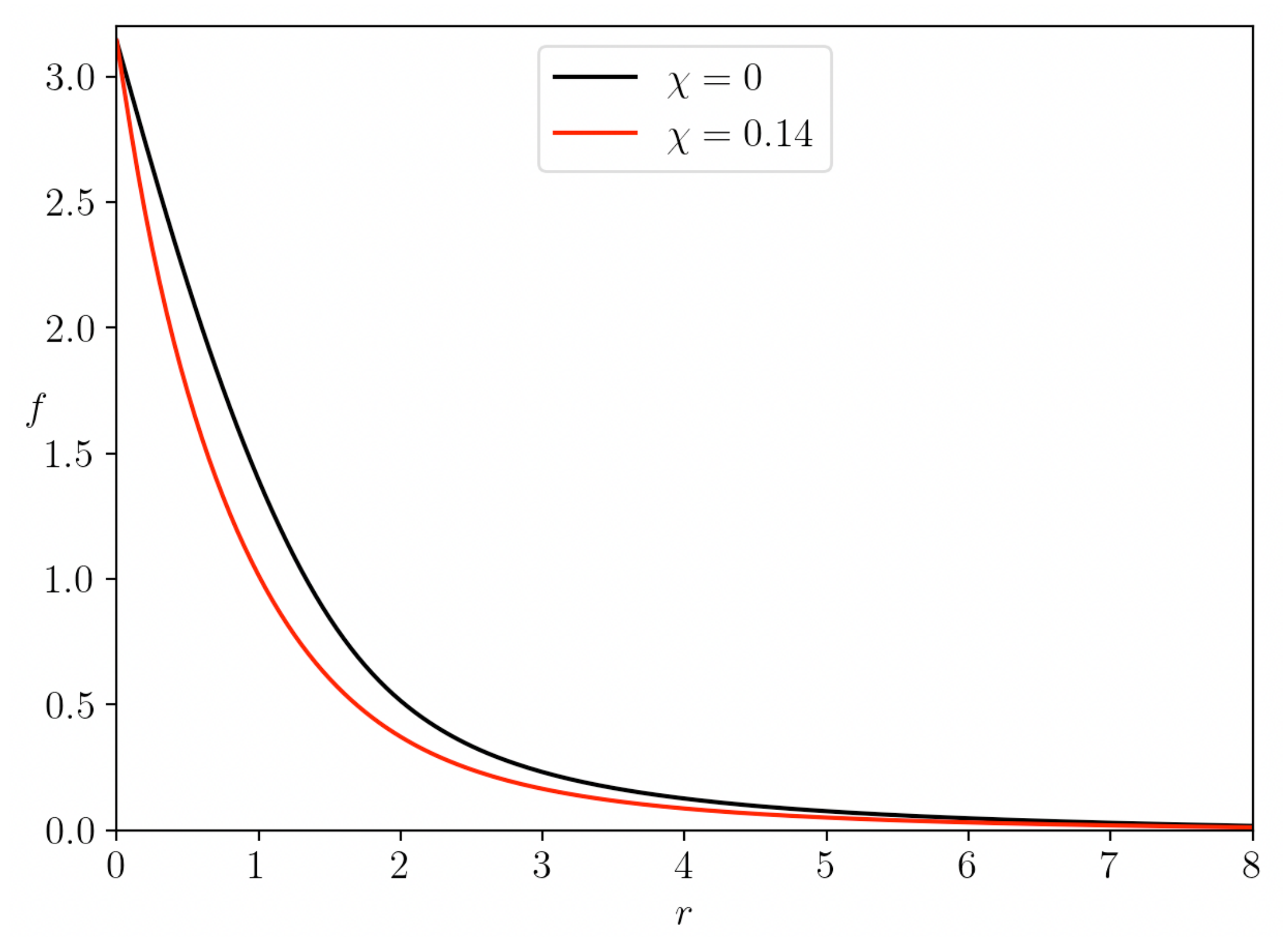}
    \caption{Left: The ratio of the single Skyrmion energy to the bound, as a function of the symmetry breaking parameter $\chi$. Right: The profile function $f(r)$ for $\chi=0$ and $\chi=0.14$.}
    \label{chi1}\end{figure}

To study the single Skyrmion in the theory~\eqref{csb}, apply the hedgehog ansatz
\begin{equation}
\sigma =\cos f, \qquad \pi_i=\frac{x_i}{r} \sin f,
\label{hh}
\end{equation}
with $f(r)$ being a monotonic decreasing profile function that satisfies the boundary conditions $f(0)=\pi$ and $f(\infty)=0.$ This yields a spherically symmetric energy density with the energy given by
\begin{gather}
E=4\pi \int_0^\infty\bigg\{
r^2{f'}^2
+2\big({f'}^2+1\big)\sin^2 f+\frac{\sin^4 f}{r^2}
\nonumber\\
\hphantom{E=4\pi \int}{}-8\chi\bigg({f'}^2(1-3\cos f)-\frac{2}{r^2}\cos f\sin^2f\bigg)\sin f\sqrt{1+\cos f}\nonumber\\
\hphantom{E=4\pi \int}{}+\frac{8}{5}\chi^2(1+\cos f)\bigg({f'}^2\big(41\cos^2f-30\cos f+9\big)
+8\frac{\sin^2f}{r^2}\big(7\cos^2f-2+2r^2\big)\bigg)\nonumber\\
\hphantom{E=4\pi \int}{}-\frac{1024}{5r^2}\chi^3(1+\cos f)^{3/2}\sin^3f\cos f
+\frac{6144}{35r^2}\chi^4(1+\cos f)^2\sin^4f
\bigg\} {\rm d}r.
\label{pro1}
\end{gather}
For each value of $\chi$ the energy is obtained by calculating the energy minimizing profile function numerically using an annealing algorithm. The result is displayed in the left plot in Figure~\ref{chi1}, where the ratio of the energy to the bound is shown as a function of the symmetry breaking parameter $\chi.$ It can be seen that as $\chi$ increases from zero, the excess energy initially decreases from around $23\%$ in the standard Skyrme model to around $7\%$ for a critical value of $\chi=0.14$ (to two decimal places) and then begins to increase again. Recall from an earlier comment that this single Skyrmion excess energy must be greater than $6\%$, as it cannot be lower than the value obtained in the $\rho$ meson theory that allows an arbitrary unconstrained vector field $V_i$ in place of the specific form used in~\eqref{ansatz}.

The energy minimizing profile function $f(r)$ is displayed in the right plot in Figure~\ref{chi1} for the standard Skyrme model ($\chi=0$, upper curve) and at the critical value ($\chi=0.14$, lower curve). It can be seen that the change in the profile function is rather mild, with the Skyrmion being a~little smaller in the modified theory.

The fact that the energy at the critical value is close to the value in the $\rho$ meson theory reveals that the expression used in~\eqref{ansatz} is indeed a good surrogate for the inclusion of $\rho$ mesons and justifies this choice. It also implies that the new terms in~\eqref{csb} may be thought of in a~similar vein to terms that would be generated by integrating out the $\rho$ mesons in the theory with pions and $\rho$ mesons, to leave an effective theory with just pions. A related point of view is that the~$\rho$ mesons have been treated as excitations of pions, extending the philosophy that all particles, mesons and baryons, are modelled as excitations of the pion field.

In the standard Skyrme model the minimal energy multi-Skyrmions can be approximated with a good accuracy using the rational map approximation~\cite{Houghton:1997kg}. It therefore seems reasonable to suppose that this approximation is also valid in the modified model, at least for small values of~$\chi.$ This approximation uses spherical polar coordinates, $r$, $\theta$, $\varphi$, with Riemann sphere coordinate $z=\tan(\theta/2){\rm e}^{{\rm i}\varphi}.$ The input to create a charge $B$ Skyrmion is $W(z)$, a degree $B$ rational map in~$z$. For $B=1,2,3,4$ these maps are given by
\begin{equation}
W=z, \qquad
W=z^2, \qquad
W=\frac{z^3-\sqrt{3}{\rm i}z}{\sqrt{3}{\rm i}z^2-1},\qquad
W=\frac{z^4+2\sqrt{3}{\rm i}z^2+1}{z^4-2\sqrt{3}{\rm i}z^2+1},
\label{ratmaps}
\end{equation}
with the property that for each degree they minimize the angular integral
\[
   {\cal I}=\frac{1}{4\pi}\int \bigg(\frac{\big(1+|z|^2\big)|W'|}{1+|W|^2}\bigg)^4 \frac{2{\rm i}\,{\rm d}z{\rm d}\bar z}{(1+|z|^2)^2},
   \]
   with values given by $1.0$, $5.8$, $13.6$, $20.7$, respectively.

   The rational map approximation is a generalization of the hedgehog ansatz~\eqref{hh} to
   \begin{equation}
   \sigma =\cos f, \qquad (\pi_1,\pi_2,\pi_3)=\frac{\sin f}{1+|W|^2}\big(W+\overline {W},{\rm i}(\overline{W}-W),1-|W|^2\big),
   \label{rma}
   \end{equation}
   and indeed is identical to the hedgehog ansatz for $W=z.$ Substituting~\eqref{rma} into the energy~\eqref{csb} yields
   \begin{gather*}
E=4\pi \int_0^\infty\bigg\{
r^2{f'}^2
+2B\big({f'}^2+1\big)\sin^2 f+\frac{{\cal I}\sin^4 f}{r^2}
\nonumber\\
\hphantom{E=}{}-8\chi\bigg(B{f'}^2(1-3\cos f)-\frac{2{\cal I}}{r^2}\cos f\sin^2f\bigg)\sin f\sqrt{1+\cos f}\nonumber\\
\hphantom{E=}{}+\frac{8}{5}\chi^2(1+\cos f)\bigg(B{f'}^2\big(41\cos^2f-30\cos f+9\big)
+8\frac{\sin^2f}{r^2}\big({\cal I}\big(7\cos^2f-2\big)+2Br^2\big)\bigg)\nonumber\\
\hphantom{E=}{}-\frac{1024}{5r^2}\chi^3{\cal I}(1+\cos f)^{3/2}\sin^3f\cos f
+\frac{6144}{35r^2}\chi^4{\cal I}(1+\cos f)^2\sin^4f
 \bigg\} {\rm d}r,
\end{gather*}
which is a generalization of~\eqref{pro1}, with the two being identical for $B={\cal I}=1.$

\begin{figure}[t]\centering
    \includegraphics[width=0.46\columnwidth]{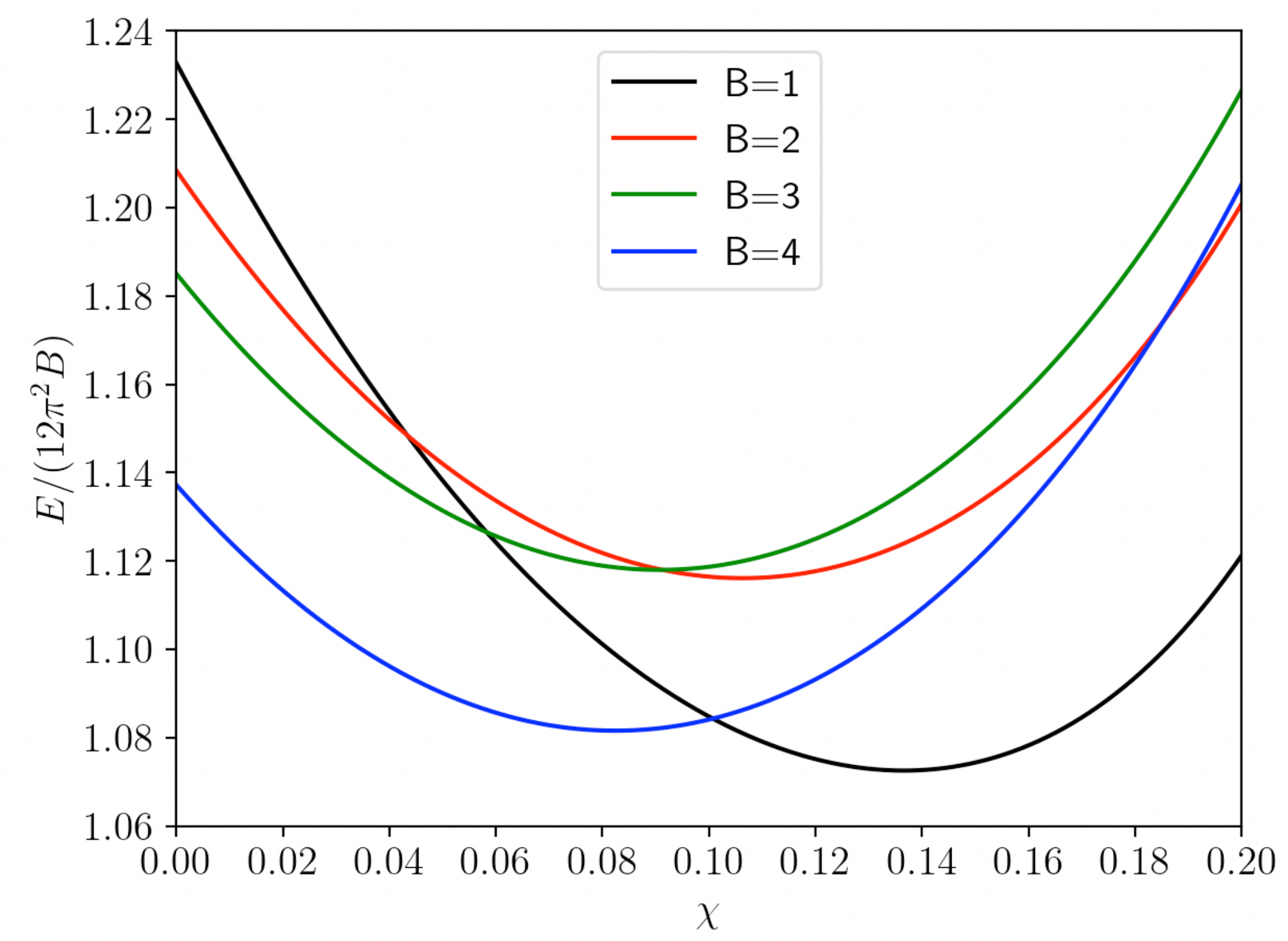} \qquad
      \includegraphics[width=0.46\columnwidth]{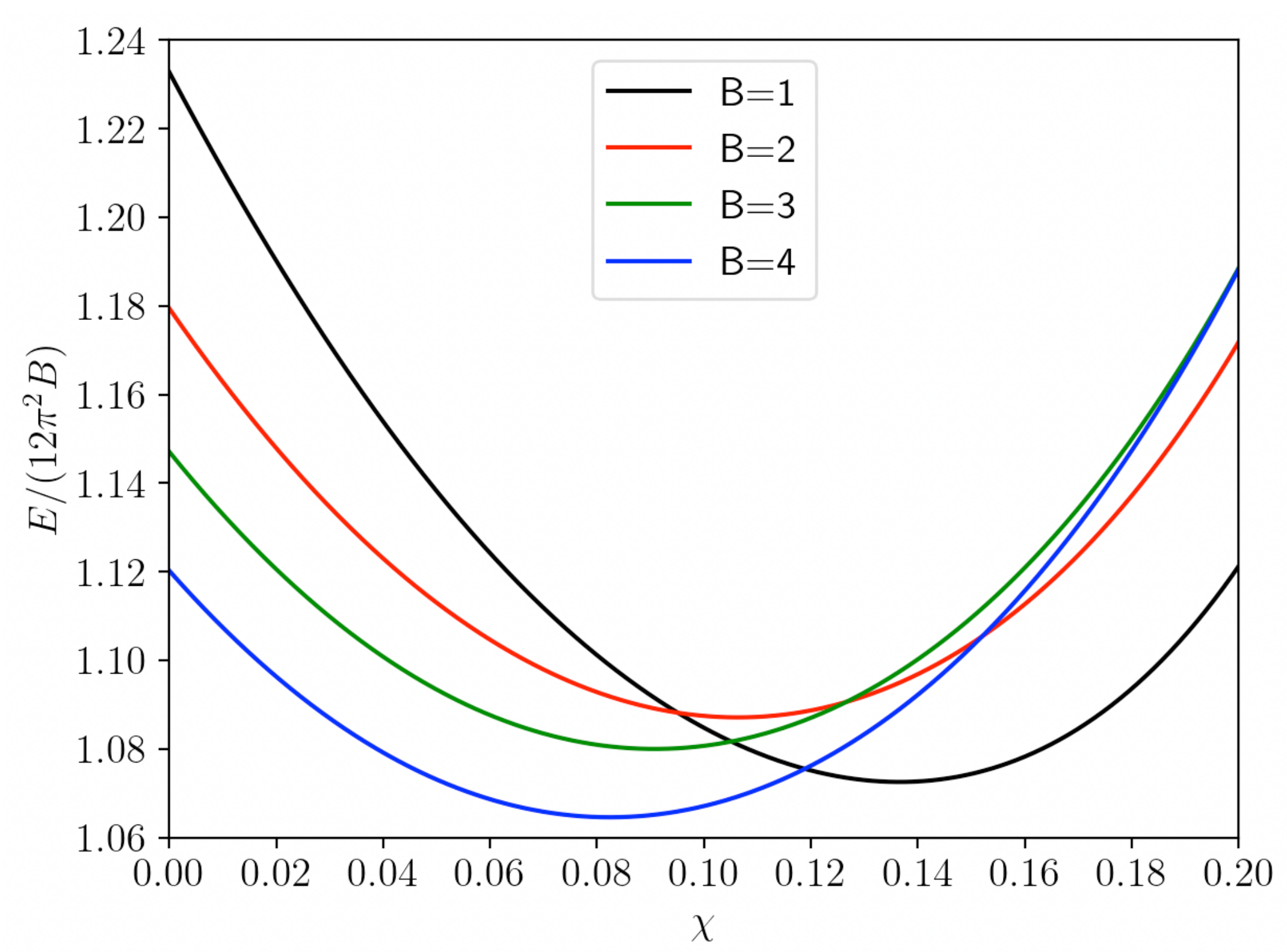}
    \caption{Left: The ratio of the rational map Skyrmion energies to the bound, as a function of the symmetry breaking parameter $\chi$, for baryon numbers one to four. Right: A similar plot but using the estimated Skyrmion energies that are exact when $\chi=0.$}
    \label{chi1234}\end{figure}
For each value of $\chi$ an annealing algorithm is applied to calculate the energy minimizing profile functions for $B=1,2,3,4$, and the resulting ratios of the energies to the bound $12\pi^2 B$ are plotted in the left image in Figure~\ref{chi1234}. Note that the points at which the $B=1$ curve crosses the other curves should not be interpreted as indicating that the corresponding charge~$B$ Skyrmions are unstable, because for $B>1$ the rational map approximation only provides an upper bound on the Skyrmion energy. Indeed, in the standard Skyrme model ($\chi=0$) numerical field theory calculations~\cite{Battye:2001qn} reveal that the rational map approximation typically overestimates the energy by something in the region of one to two percent, but for low charges this can be over three percent. Such overestimates take on more significance in modified models, as they are not small compared to binding energies.

An elementary attempt to address this overestimate issue is to use the previous numerical field theory computations at $\chi=0$ to remove the overestimate in the rational map approximation at this value, so that the intercepts of the energy curves are correct and should therefore provide better estimates of the Skyrmion energies for small $\chi.$ The result is displayed in the right image in Figure~\ref{chi1234}. This is encouraging, as the $B=1$ curve now crosses the other curves in a small interval of $\chi$, suggesting that it may be possible to have small binding energies for all $B$ for a~suitable value of $\chi.$ A caveat to this result is the assumption that the symmetries given by the rational maps~\eqref{ratmaps}, namely axial, tetrahedral and cubic, for $B=2,3,4$ respectively, persist as symmetries of the minimal energy Skyrmions when $\chi\ne 0.$ This could be violated, for example by the $B=2$ Skyrmion splitting into a pair of distinct single Skyrmions, and this could lower the energy below the estimate presented in this plot. To investigate such a possibility requires performing full numerical field theory simulations, which is a significant task, but the results presented here provide evidence that this would be a worthwhile endeavour for the future.

\section{Conclusion}

By exploiting a connection between the Skyrme model and Yang--Mills theory with an extra dimension, a variant of the Skyrme model has been presented in which derivative terms break chiral symmetry to isospin symmetry. The new terms are controlled by a single parameter $\chi$ and vanish when $\chi=0$. The main motivation for the new variant is the fact that $\chi$ can be chosen so that the energy of a single Skyrmion is much closer to the topological energy bound than when $\chi=0$. This is encouraging for obtaining more realistic binding energies, with an analysis using the rational map approximation producing promising initial results. Further investigations will require numerical field theory computations and the addition of a pion mass term. It is expected that the structure of Skyrmions in this new variant of the Skyrme model will be more sensitive to the pion mass for values of $\chi$ that yield low binding energies, as this is the case in the model with $\rho$ mesons that this new theory aims to mimic.

\pdfbookmark[1]{References}{ref}
\LastPageEnding

\end{document}